\documentclass[comsoc, journal]{IEEEtran}
\usepackage[utf8]{inputenc}
\usepackage[T1]{fontenc}

\usepackage{etoolbox}
\usepackage{xstring}
\usepackage{balance}
\DeclareListParser{\doslashlist}{/}
\newcounter{ndnNameComponentCounter}%
\newcommand{\name}[1]{{%
  \setcounter{ndnNameComponentCounter}{0}%
  \renewcommand{\do}[1]{{%
    \ifnumgreater{\value{ndnNameComponentCounter}}{0}{\allowbreak/}{}%
    \ifnumodd{\value{ndnNameComponentCounter}}{}{}%
    ##1}%
    \stepcounter{ndnNameComponentCounter}}%
``{\fontfamily{cmtt}\small\selectfont\IfBeginWith{#1}{/}{/}{}\doslashlist{#1}}''%
}}

\usepackage[utf8]{inputenc}
\usepackage{graphicx}
\usepackage[T1]{fontenc}
\usepackage{subcaption}
\usepackage{color,soul}
\usepackage{xspace}
\usepackage{enumitem}
\usepackage{titlecaps}
\usepackage{tabularx}
\usepackage{balance}
\usepackage{subcaption}
\usepackage{float}
\usepackage{url}

\newcommand{\sol}{{\em NDNTP}\xspace}

\newcommand{\ie}{{\em i.e.,}\ }
\newcommand{\eg}{{\em e.g.,}\ }

\newcommand{\spyros}{\color{black}}
\newcommand{\abde}{\color{black}}

\title{\sol: A Named Data {\spyros Networking} Time Protocol}

\author{
	Abderrahmen Mtibaa,
	and Spyridon Mastorakis
	
	\thanks{A. Mtibaa (corresponding author) is with the University of Missouri-St. Louis, US.}
	
	\thanks{S. Mastorakis is with the University of Nebraska, Omaha, US.}
}

\makeatletter
\def\ps@IEEEtitlepagestyle{%
  \def\@oddfoot{\mycopyrightnotice}%
  \def\@evenfoot{}%
}
\def\mycopyrightnotice{%
  {\footnotesize  This paper has been accepted for publication by the IEEE Network Magazine.\hfill}
  \gdef\mycopyrightnotice{}
}

\begin{document}

\maketitle

\begin{abstract}

Named Data Networking (NDN) architectural features, including multicast data delivery, stateful forwarding, and in-network data caching, have shown promise for applications such as video streaming and file sharing.
However, collaborative applications, requiring a multi-producer participation introduce new NDN design challenges. In this paper, we highlight these challenges in the context of the Network Time Protocol (NTP) and one of its most widely-used deployments for NTP server discovery, the NTP pool project. 
We discuss the design requirements for the support of NTP and NTP pool and present general directions for the design of a time synchronization protocol over NDN, coined Named Data Networking Time Protocol (\sol). 

\vspace{-0.5cm}


\end{abstract}
\IEEEpeerreviewmaketitle

\section {Introduction}

Information-Centric Networking (ICN)~\cite{xylomenos2013survey} {\spyros and a prominent} realization of its vision, the Named Data Networking (NDN) architecture~\cite{zhang2014named}, use application-defined names to retrieve secured data through a stateful forwarding plane. This waives the need for costly mappings between the application data and its physical location(s). 
The NDN design and its built-in features, {\spyros such as multicast data delivery and in-network caching,} have shown promise for applications and services such as video streaming, file sharing, and others~\cite{mastorakis2017ntorrent, gusev2015ndn}.


The ICN/NDN community has worked on adapting IP-based {\spyros applications and services} to run on top of NDN~\cite{mastorakis2017ntorrent}, including general guidelines on adapting non-NDN applications {\spyros to function over NDN}~\cite{liang2018ndnizing}. However, the NDN adoption {\spyros for certain categories of applications} remains largely unexplored. For instance, crowd-sensing and crowd-sourcing {\spyros applications and services that enable multiple producers to return different content for the same Interest may introduce new challenges to the existing} NDN design.



In this paper, we consider the Network Time Protocol (NTP)~\cite{mills1991internet} as a use-case {\spyros given the universal need of applications and devices for a time synchronization service. Through this use-case, we highlight new NDN design requirements} and design an NDN-based service, {\spyros called Named Data Networking Time Protocol (\sol)}, to provide time synchronization to applications, systems, and devices {\spyros within islands of NDN connectivity}. We present a preliminary investigation of how functionality equivalent to the IP-based NTP can be provided over NDN. We further consider NTP pool~\cite{ntp-pool} as one of the most widely-used NTP server discovery deployments. Motivated by the requirements of NTP and NTP pool, {\spyros we discuss the \sol design as well as} enhancements of the native NDN architectural features to support the NTP functionality. 

{\spyros As part of our work, we} propose design directions for: (i) forwarding time synchronization requests towards different sets of servers and retrieving multiple time samples from each of these servers; (ii) fine-grained control over how far in the network time synchronization requests can travel; and (iii) controlling the NDN in-network caching and request aggregation mechanisms to enable the retrieval of up-to-date time samples. The design directions discussed throughout the paper can be leveraged {\spyros by various NDN} applications that share the same requirements as \sol.

The rest of this paper is organized as follows: Section~\ref{sec:back} presents {\spyros a brief background on NDN, NTP, and NTP pool}, Section~\ref{sec:design} discusses the design requirements {\spyros of \sol}, while Sections~\ref{sec:multi-server},~\ref{sec:prob}, and~\ref{sec:others} {\spyros present} alternative directions to satisfy each of the design requirement. Section~\ref{sec:disc} briefly discusses \sol extensions, while Section~\ref{sec:concl} concludes the paper. 




\section {{\spyros Background}}
\label{sec:back}

{\spyros In this section, we first present an NDN primer. Then we present some background on NTP and NTP pool to prepare readers for the discussion in the rest of the paper.}

\subsection{\spyros {NDN Background}}

{\spyros NDN~\cite{zhang2014named} features a data-centric, receiver-driven communication model, where each piece of data has an application-defined name. \emph {Consumer applications} send requests for named data, called Interests, towards \emph{data producer applications}. Each Interest carries the name of the requested data. Once a data producer receives an Interest, it sends back a Data packet, which is cryptographically signed by the data producer and contains the requested data.

Interests are forwarded based on their names towards producers by NDN forwarders. To achieve that, NDN forwarders make use of a Forwarding Information Base (FIB), which contains name prefixes along with a set of outgoing interfaces for each prefix. Forwarders also maintain a Pending Interest Table (PIT), where they store network state for each Interest that has been forwarded, but the corresponding data has not been received yet. A Data packet follows the reverse path of the corresponding Interest based on the network state in PIT and can be forwarded all the way back to a consumer only if there is a valid PIT entry for the corresponding Interest at each hop forwarder. Once a Data packet is retrieved in response to an Interest, the corresponding PIT entry at each hop forwarder will be consumed. This happens as part of the ``flow balance'' principle of NDN, which mandates that each Interest packet can bring back only one Data packet from each hop forwarder. If a PIT entry stays open to allow for more than one Data packets to be returned to a consumer, the ``flow balance'' principle is violated. Finally, each forwarder is equipped with a Contest Store (CS), where recently received Data packets are cached to satisfy Interests for the same data in the future.}




\subsection {{\spyros NTP and NTP Pool Background}}

\begin{figure}[t]
    \centering
    \includegraphics[width=0.9\linewidth]{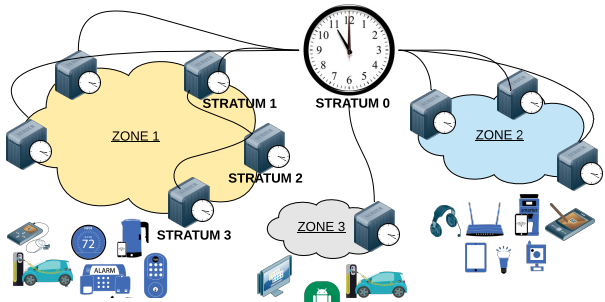}
    \vspace{-0.2cm}
    \caption{Zones of NTP servers based on their geolocations in NTP pool; clients reach servers in the same or a nearby zone.}
    \label{fig:ntp}\vspace{-0.5cm}
\end{figure}

NTP~\cite{mills1991internet} has been one of the longest running protocols on the Internet. It was created due to the need of {\abde applications and services} for time synchronization over packet-switched networks. Discovering NTP servers and automating this process for clients has been essential for the deployment of NTP. {\abde Among the different service discovery schemes proposed (\eg broadcast/multicast/manycast servers)}, NTP pool has been one of the most widely-used ones~\cite{ntp-pool}. NTP pool is a cluster of volunteer NTP servers used by large numbers of clients around the world {\spyros (\eg it has been the default time synchronization option for most Linux distributions)}. NTP pool groups NTP servers based on their IP address geolocations into continental and country zones ({\spyros Fig.~\ref{fig:ntp}}).

{\spyros Each NTP server} in the pool is assigned a score, which reflects {\abde the accuracy of the provided time samples are} 
This score is determined by a monitoring station. Once a new server joins the pool, it is assigned a low score by default. The monitoring station probes {\abde this newly joined} (and every other) server over time, verifying the accuracy of its clock. As servers respond with accurate clock readings to the requests of the monitoring station, {\abde their scores improve}. {\abde Once the score of a server reaches a certain threshold}, the server will start receiving requests by NTP clients. NTP clients discover servers in the pool by querying DNS for the \name{pool.ntp.org} domain name. The DNS resolution of \name{pool.ntp.org} would usually return servers within or close to the client's country.


\section{Design Requirements of \sol}
\label{sec:design}

In this section, we first give an overview of the \sol network model. {\abde Then, we present \sol's design goals and requirements}, focusing on the NDN features that need to be enhanced in the context of this design.

\subsection{Network Model and Assumptions}

We consider NDN islands, where consumers (\sol clients) are scattered geographically into multiple zones {\spyros such as} Europe, America, and Asia. In each zone, there is a number of \sol servers acting as NDN data producers. \sol clients are inter-connected with \sol servers through one or more network hops (NDN forwarders). 


The \sol servers receive time requests (Interest packets) under the \name{/NDNTP/time} namespace sent by \sol clients, who initiate the time synchronization process. Servers respond to requests by generating and signing \sol responses ({\abde Data packets}), which contain the current timestamp and other necessary fields as determined by the NTP specification~\cite{mills2010network}. To receive requests from clients, servers announce the \name{/NDNTP/time} namespace to the network. {\spyros A name-based routing protocol, such as NLSR~\cite{hoque2013nlsr}),} propagates the announcement and establishes routes towards servers (\ie install the proper forwarding information on the FIB of NDN forwarders). 



\subsection{\sol's Design {\abde Requirements}}

Deploying \sol in an NDN island can be challenging, since functionality requirements motivated by NTP and NTP pool require the augmentation of certain NDN architectural features. In the rest of this section, we present the main functionality requirements of the \sol design and motivate the challenges on fulfilling them over NDN.



\begin{figure}[t]
\centering
\includegraphics[width=1.0\linewidth]{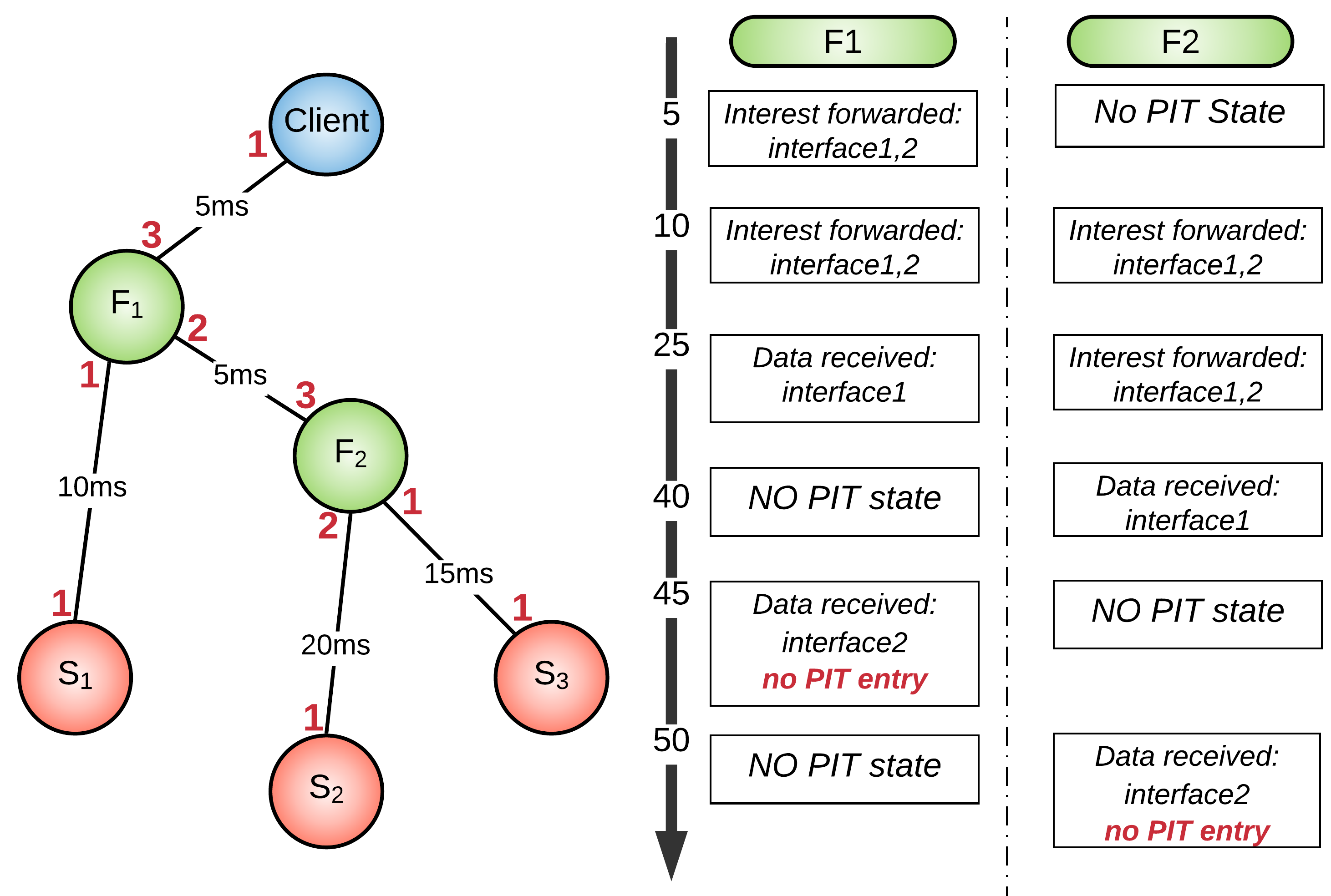}
\vspace{-0.5cm}
\caption{Challenge to collect {\spyros multiple Data packets for a single multicast Interest. The client} sends an Interest that reaches all servers $S_1, S_2, S_3$. Link delays indicate that $S_1$ will receive and send its data back earlier than the other servers.}
\label{figure:1interestMdata}\vspace{-0.4cm}
\end{figure}

\noindent \textbf{Multi-server time sample fetching:} The best current practices of NTP deployment~\cite{reilly2017network} indicate that a client should contact more than one NTP servers in order to select accurate time sources and disregard unreliable ones. NDN inherently facilitates reaching multiple servers through its multipath/multicast nature. {\spyros In this context, clients} send requests for \name{/NDNTP/time}, which will be satisfied by any server running the time synchronization service. 
{\spyros In \sol,} we would like multicast requests {\spyros in order to bring a response} from each server they reached back to a client. However, NDN allows only a {\spyros single Data packet} to use the reverse path to the client--often the closest server will use the {\spyros network state stored in PIT}, thus responses sent by other servers will not find PIT state to be returned to the client. {\spyros Fig.~\ref{figure:1interestMdata} illustrates such a scenario, where a client} reaching three different servers would receive a {\spyros single Data packet} from the closest server (\ie $S_1$). This packet will consume the {\spyros PIT entry at forwarder} $F_1$, leaving data from other servers without a path back to the client.

Moreover, clients typically request multiple time samples from the same set of servers to increase the accuracy of the time synchronization process. In NDN, clients rely on the network to guide requests towards the ``best'' (\ie often the closest) server, without being able to control whether subsequent time synchronization requests will reach the same or different servers. {\spyros To this end}, additional mechanisms are needed to let more than one requests be forwarded towards the same set of servers. All these challenges are discussed and different design directions are proposed in Section~\ref{sec:multi-server}.


\noindent \textbf{Distance-based server selection:} Through the NTP pool, clients often select a set of NTP servers that are in the same country or continental zone as {\spyros themselves.} 
Clients discover different sets of servers over time in order to: (i) avoid the impact of malicious groups of neighboring servers colluding to return bogus timestamps to clients
; and (ii) balance the load among servers in a zone. 

{\spyros NDN makes use of application-defined names for communication purposes}, enabling access to services typically from the closest source/server that can offer them. However, to achieve distance-based server selection, \sol clients may need mechanisms for fine-grained control over how far their requests should travel and how these requests can be satisfied by servers that are potentially not the closest ones to clients. We further discuss this challenge and propose directions to address {\spyros it by enhancing the NDN} design in Section~\ref{sec:prob}.

\noindent \textbf{Time synchronization freshness and accuracy:} NTP clients aim to retrieve fresh (up-to-date) time samples, increasing the accuracy of the time synchronization process. However, due to NDN in-network caching, \sol client requests for time synchronization can be satisfied with outdated responses that have been cached in the network. Moreover, \sol requests (from the same or different clients) reaching a forwarder that has another \sol request pending in its PIT could result in request aggregation. That is, subsequent \sol requests may not be forwarded to a server, but will be satisfied when the response to the first pending request is received by the forwarder. This could skew {\spyros the client round-trip delay measurements, impacting} the accuracy of the time synchronization process. We further discuss these challenges and propose directions to tackle them in Section~\ref{sec:others}.






\section{Enabling Multi-Server Time Sample Fetching}
\label{sec:multi-server}

In Section~\ref{sec:design}, {\abde we present different design directions for the following \sol design goals/objectives: (i) ensure that clients can reach multiple servers and collect multiple time samples for each of these servers; and (ii) enable a multicast request to bring a response from each server that it reached back to a client}. 



\subsection{Gathering Multiple Samples From Multiple Servers}
\label{sec:unicast}
 
It is desirable for NTP clients to: (i) contact more than one NTP servers in order to {\spyros identify accurate time sources and disregard inaccurate ones}; and (ii) collect multiple time samples from each NTP server to enhance the accuracy of the synchronization process due to the dynamic nature of network conditions. The NDN communication model is inherently multipath/multicast supporting requirement (i), since an Interest can be forwarded towards multiple producers (\sol servers in our case). However, deterministically contacting the same server multiple times in the context of requirement (ii) is counter-intuitive to the purpose of NDN, where data can be retrieved from any party that can provide it. {\spyros As a result,} the NDN architecture itself does not {\spyros provide explicit mechanisms to ensure} that Interests will be forwarded along a specific path towards a certain server. Below, we present a wide spectrum of design directions to fulfill these requirements ranging from solutions of unicast nature to solutions of multicast nature.
 
The first design direction is inspired by source routing. {\spyros Clients instruct the network about which paths their Interests should take through techniques to create path labels} for Interest forwarding~\cite{moiseenko2017path}. Each Interest carries such a label, which determines the next-hop that the Interest should be forwarded to. At each hop, NDN forwarders use this label for Interest forwarding bypassing the FIB lookup process. At the beginning of their operation, \sol clients will perform a path discovery process, where they will acquire labels for paths to different servers. This process will end when a client acquires labels for the desired number of servers. The client uses the same path label to collect different time samples from the same server, since Interests carrying the same label will be forwarded {\spyros along the same path towards the same server.} 

\begin{figure}[t]
\centering
\includegraphics[width=0.9\linewidth]{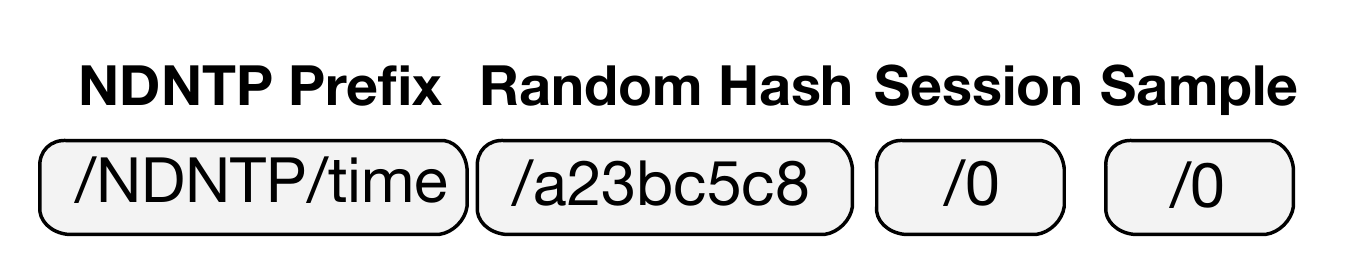}
\vspace{-0.2cm}\caption{Namespace design {\abde enabling clients to: (i) reach multiple \sol servers; and (ii) collect multiple time samples from a specific \sol server}}
\label{figure:namespace}\vspace{-0.3cm}
\end{figure}

Another direction would be to utilize the stateful NDN forwarding plane. 
Forwarding modules (strategies) can be created and deployed on NDN forwarders, so that clients can discover multiple \sol servers and collect a number of time samples from each of these servers.
Specifically, \sol clients create a ``session-like'' paradigm with a server, so that they request multiple time samples from it. An example namespace to achieve that is illustrated in {\spyros Fig.~\ref{figure:namespace}}. The name prefix includes {\spyros the \sol name prefix} (\name{/NDNTP/time}), followed by: (i) \emph{a random hash} to by-pass the aggregation of requests from different clients as further discussed in Section~\ref{sec:others} (each different hash identifies a set of sessions initiated by a client--each session is with a different server); (ii) \emph{a session number} that identifies a specific session with {\spyros a server}; and (iii) \emph{a sample number} that identifies a specific time sample requested from a server.

When an NDN forwarder receives an \sol request, it dispatches it to the corresponding forwarding {\spyros module, which identifies} requests from the same client but from different sessions. To this end, time sample requests for the same session number are forwarded along the same path, so that they reach the same \sol server. At the same time, the stateful network forwards requests for different session numbers along different paths towards different servers. 

\noindent{\abde\bf Multicast Support:} {\abde We note that both solutions mentioned above} can be extended to support multicast Interest forwarding. For instance, during the discovery process of paths performed by clients, the path labels {\spyros can include multiple next hops at each} forwarder. As a result, client Interests will be forwarded along different paths, potentially reaching different servers. In a similar manner, {\spyros in the second approach,} client Interests can include a list of session numbers instead of a single session number in order to enable multipath stateful forwarding. Clients will re-use the same list of sessions to ensure that subsequent requests for time samples reach the same set of servers. {\spyros However, directions of multicast} nature require multiple responses (one from each server that received a request) to be forwarded back to clients. We discuss solutions to that below.

\subsection{Receiving Multiple Responses Upon Sending Multicast Requests}





{\abde To address this challenge}, one direction would be to enhance the NDN communication model, so that forwarders can accept multiple {\spyros Data packets} (one per interface that the corresponding Interest was forwarded through). These {\spyros Data packets} will be aggregated into a single packet, which will be forwarded back to the forwarder's previous hop~\cite{mastorakis2020icedge}. In this way, forwarders can multicast \sol requests towards multiple servers, receive a response from each of these servers, and aggregate the responses into a single response. This ``aggregated'' response will eventually be received by the requesting \sol client. {\spyros This enhancement does not invalidate NDN's} principle of ``flow balance, {\spyros since a single response will be returned to clients.} However, it can take arbitrarily long for a forwarder to collect multiple responses through different paths, since the length of each path and the network {\spyros conditions might differ.} This could impede {\spyros the time synchronization accuracy} by skewing the round-trip delay measurements of clients.

An alternative direction consists of consuming a PIT entry only after all the expected responses have been received. Specifically, a PIT entry will stay alive and bring back one response for each outgoing interface that the corresponding \sol request was forwarded through. This would help clients to collect accurate round-trip delay measurements and improve the time synchronization accuracy. However, this direction would invalidate {\spyros the NDN ``flow balance'' principle, since corresponding PIT entries may stay open and accept several Interests.}


\section{Distance-Based Server Selection}
\label{sec:prob}

While NDN inherently supports reaching the closest server {\abde offering the time synchronization service, there might be cases where} reaching distant servers within a zone may be desirable 
in order to: (i) reduce targeted attacks, where entities controlling a group of neighboring servers collude to return bogus timestamps; and (ii) enable load balancing among nearby and distant servers. Reaching servers within similar distances from a client may also be desirable in order to reduce jitter in round-trip delays. 

\noindent{\bf Preliminary Experiments:}
To investigate whether NTP pool shares the above objectives and better understand how it operates, we performed a set of experiments. Specifically, we deployed NTP clients--two in North America (in Missouri, US {\spyros and in New} Mexico, US) and two in Europe (in Zurich, Switzerland {\spyros and in Athens,} Greece)--and we configured them to use NTP pool for time synchronization. Each client queries four NTP pool servers and collects four time samples from each of these servers (\ie the default {\spyros configuration options} for NTP) every minute over a 24-hour period\footnote{The datasets consisting of the raw timestamps and servers probed are available at \url{https://github.com/amtibaa-cmu/NTP-Data-Traces}.}.



\begin{figure}
    \centering
    \includegraphics[width=0.8\linewidth]{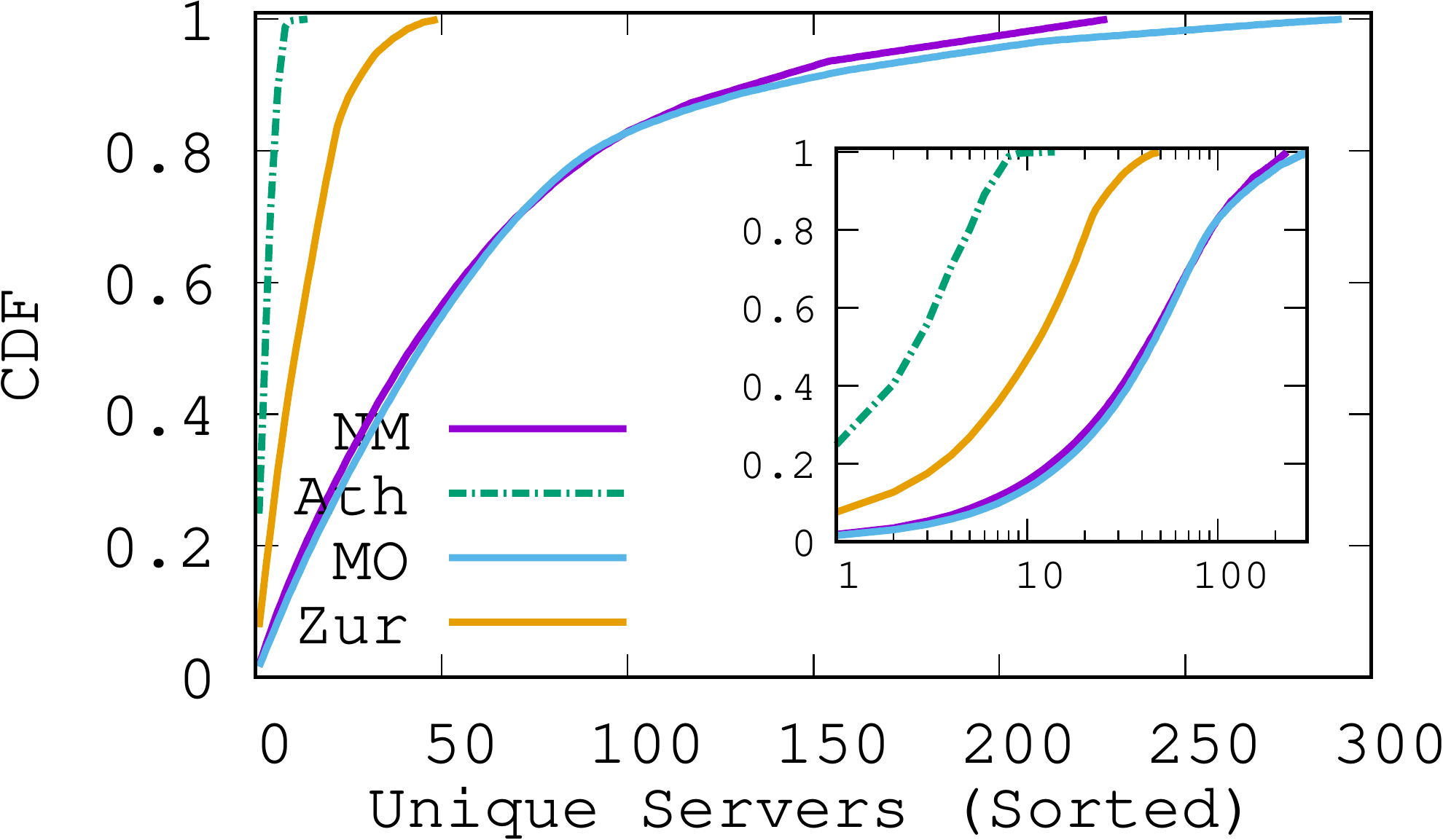}
    \vspace{-0.2cm}
\caption{Distribution of server occurrences; servers are {\spyros sorted from the most to} the least selected in the New Mexico (NM), Missouri (MO), Zurich (Zur), and Athens (Ath) traces. The inside plot is in log scale zooming into the distribution of the most popular servers.}
    \label{IP_cdf_all}\vspace{-0.4cm}
\end{figure}

{\spyros Fig.~\ref{IP_cdf_all}} presents the CDF of the unique NTP pool servers {\spyros sorted by popularity (\ie number of times a server is selected)} in the four traces (results collected from each client).
We notice that both North American traces have very similar trends characterized by almost identical CDF distributions, while the European traces have different properties compared to the traces from North America. Based on the European traces, the unique servers discovered are $15$ to $20 \times$ fewer than those discovered in North America (\eg 14 unique servers in the trace from Athens compared to 300 in the Missouri trace). Our understanding is that this is due to the difference in the zone sizes for the US and the European countries (Greece and Switzerland). For example, at the time of the experiments, there were 774 NTP servers in the zone of the US, while only 15 NTP servers in the zone of Greece. 

Moreover, we found that the NTP pool servers in Europe are closer to {\spyros the clients than the servers in the US}. For example, more than 80\% of the servers discovered from Athens are within 50 kilometers from the client, while less than 10\% of the servers discovered from New Mexico {\spyros are within 1000 kilometers from the client (\ie $250 \times$ further than the 50-kilometer} neighboring circle in Athens). This is due to the different physical sizes of the countries--in terms of landmass, the US is 9,833,000 $km^2$, while Greece is only 131,957 $km^2$.

\begin{figure}[t]
\centering
 \begin{subfigure}[b]{0.49\linewidth}
 \includegraphics[width=\textwidth]{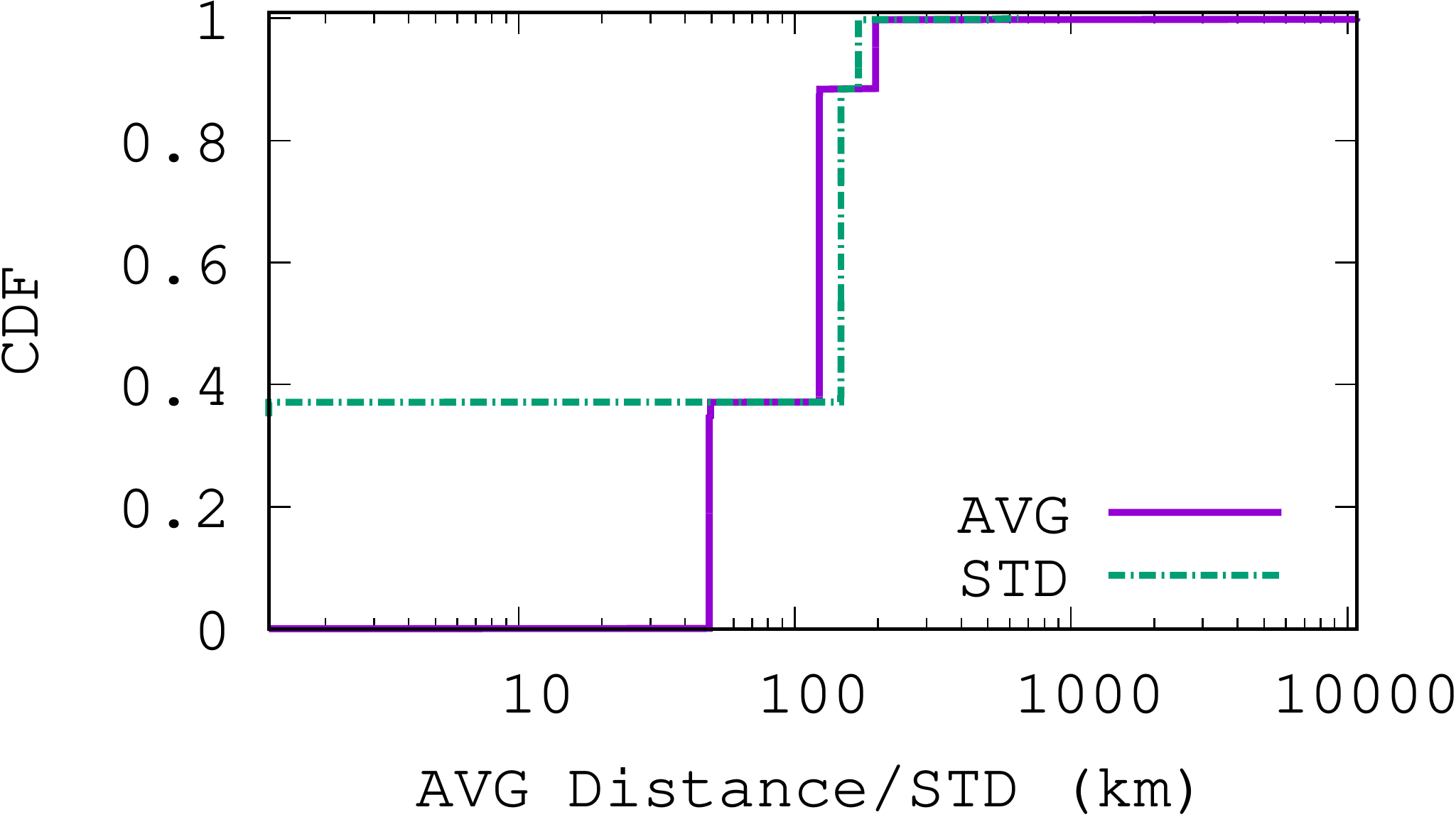}
 \caption{Athens}
 \label{dist_ath}
 \end{subfigure}
 \begin{subfigure}[b]{0.46\linewidth}
 \includegraphics[width=\textwidth]{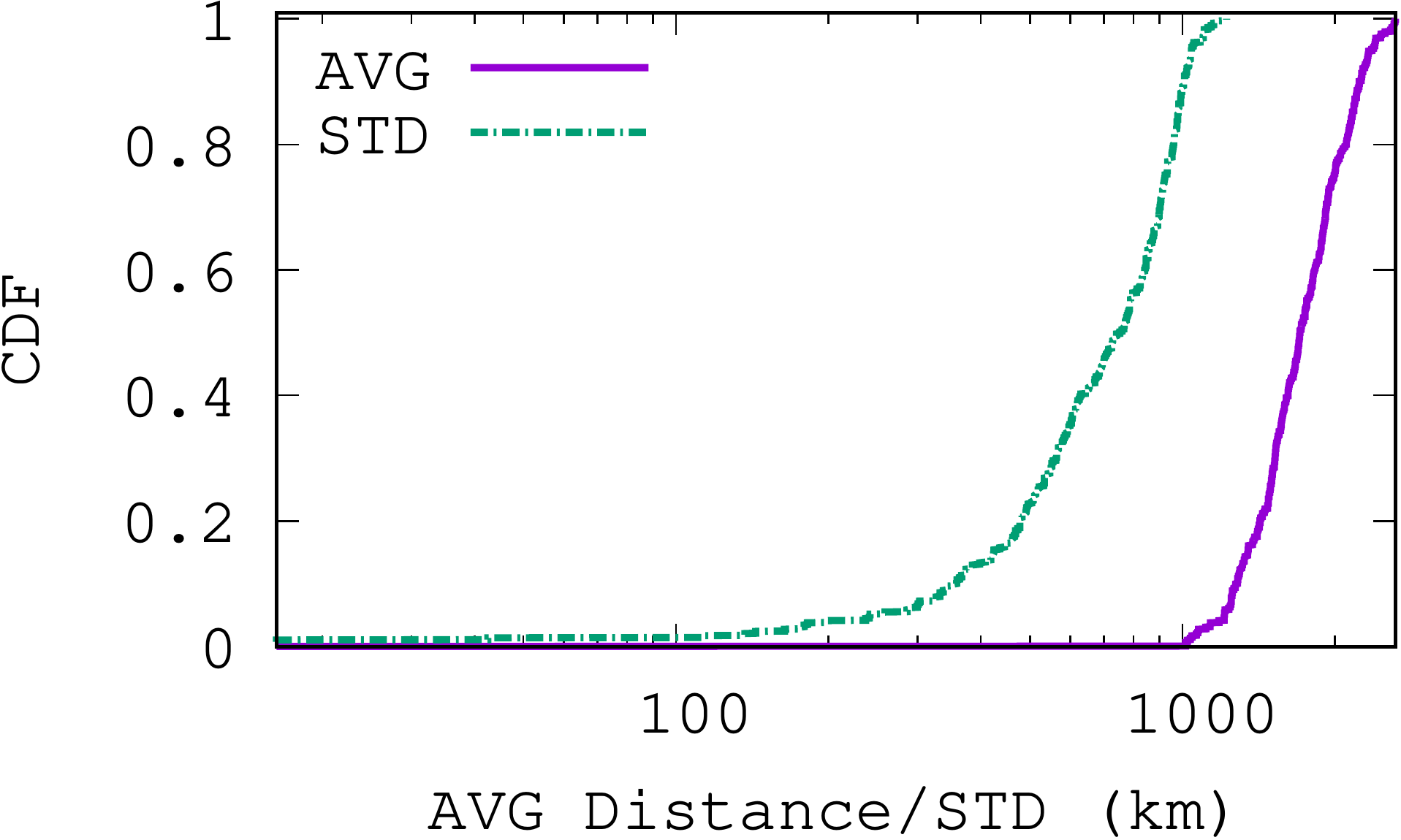}
 \caption{New Mexico}
  \label{dist_nmsu}
 \end{subfigure}
 \vspace{-0.2cm}\caption{{\abde CDF of the average distances} (AVG) and standard deviations (STD) per run {\abde for the (a) Athens and the (b) New Mexico traces}}
 \label{distance_cdf}\vspace{-0.5cm}
\end{figure}


In {\spyros Fig.~\ref{distance_cdf}}, we illustrate the average distances (in kilometers) and the {\spyros standard deviations per} run of the NTP protocol (four servers are selected and four times samples are collected from each server per run) {\spyros for the (a) Athens and the (b) New Mexico} traces respectively\footnote{For better readability, we selected one trace from Europe and another from the US. We have verified the validity of our conclusions for all collected traces.}. First, we notice that the four servers selected in each run tend to have similar distances from the client as shown by the lower values in standard deviation distributions of {\spyros Fig.~\ref{distance_cdf}}. Furthermore, we observe that in 40\% of the runs for the Athens trace, requests reached four servers within the same distance of 45 to 50 kilometers. On the other hand, the four servers selected in each run of the New Mexico trace have different distances from the client, resulting in larger standard deviation values. Our understanding is that this is due to the larger physical surface of the US compared to Greece.


\subsection{Lessons Learned} 

The lessons we learned from these experiments fall into three main categories: (i) {\em zone probing}: NTP pool employs a zone probing scheme that consists of reaching (nearby and {\spyros occasionally distant}) NTP servers from within the client's (continental and/or country) zone; (ii) {\em similar distance probing}: most servers selected in each run of the NTP protocol are within similar distances from the client, thus responses return to the client with similar delays; and (iii) {\em reaching a different set of NTP servers in each run}: NTP clients that make use of NTP pool reach, in general, a different set of servers for every run. Our data showed that the same US servers are rarely selected more {\spyros than a few times in the 24-hour period}. However, in zones, such as in Greece and Switzerland, where the number of NTP servers in the zone is small, servers are selected multiple times. In addition to that, distant servers are selected, although infrequently (as shown in {\spyros Fig.~\ref{distance_cdf}}), to reduce the reliance on a small set of servers that may form a collaborative malicious group (collusion attacks).


\subsection{{\spyros Distance-Based Server Selection in \sol}}
\label{subsec:probing}

According to the lessons we learned from our experiments, \sol needs to be able to reach servers within a given ``{\em zone}'' and within controlled distances from clients. While a zone in an IP-based NTP pool design refers to a country or a continent, which may have largely variable sizes, we define a zone as the region within a certain number of hops from a client. For instance, a client in New Mexico will be able to reach servers within $h$ hops making the case of reaching servers in a different country likely in addition to reaching servers in other US states. This can be achieved through the deployment of stateful forwarding modules that allow clients to reach not only the closest {\spyros but also distant \sol servers}. We propose two client-based directions that enable fine-grained control over how far a request can travel.

{\spyros The first design direction} consists of taking advantage of the \emph{hop limit} field in Interest packets. Clients that would like their requests to reach servers at least $h$ hops away will set the hop limit value of their requests to $h$. {\spyros Each forwarder decrements} the hop limit value of \sol Interests and forwards them through the outgoing interface with the highest cost--in NDN, outgoing interfaces for a given name prefix are associated with a cost, thus the lower the interface cost is, the closer a server that can satisfy Interests for this name prefix is. {\spyros When the hop limit value of} an Interest reaches a given threshold, the forwarding plane switches to using the outgoing interface with the lowest cost for this Interest. {\spyros As a result, after} this point, the Interest will be forwarded with the goal of reaching the closest server(s). Through this approach, all the selected servers are at least $h$ hops away and within similar distances from clients, so that round-trip delays for requests are similar.

{\spyros The second direction would be} to enable clients to include a probability to the name of their \sol requests. The lower this probability is, the higher will be the cost of the outgoing interface chosen by forwarders. For instance, a request with a name \name{/NDNTP/time/P=0.3} would be handled by forwarders so that they choose the outgoing interface with the highest cost with a probability of 0.7 and the lowest cost interface towards a server with {\spyros probability of 0.3}. This method allows clients to have fine-grained control over when requests should reach closer versus more distant servers, however, it does not ensure reaching servers at certain distances {\spyros (\eg at least $h$ hops away)} from clients. 

\section{In-Network Caching and Request Aggregation in \sol}
\label{sec:others}

In this section, we discuss how to achieve time synchronization in the face of in-network caching and request aggregation.

\noindent \textbf{In-network caching:}
To avoid fetching outdated cached responses, the requests of \sol clients need to be satisfied directly by \sol servers rather than in-network caches. To this end, servers can set the value of the {\em Freshness Period} field of the responses they generate to a reasonably low value (or even 0), so that their responses become (almost) {\spyros instantly stale when they are cached in the network}~\cite{mastorakis2018real}. Subsequently, clients send \sol Interests that contain the {\em MustBeFresh} {\spyros field to avoid retrieving} stale cached \sol responses.

An additional challenge arises when servers misbehave by ignoring this guideline and assign a large Freshness Period value to the generated responses. To this end, appropriate cache management policies can be deployed in the network to prevent forwarders from caching \sol responses {\spyros for a long period of time} or entirely avoid caching \sol responses. Note that this solution assumes trusted forwarders--we discuss security considerations and potential solutions in cases of malicious \sol servers and forwarders in Section~\ref{subsec:security}.

\noindent \textbf{\sol request aggregation in PIT:} 
\sol requests may be aggregated in PIT when a forwarder has a PIT entry for a request with the same name. {\spyros As a result, subsequent \sol requests} with the same name may not be forwarded to a server, but will be satisfied when the response to the first pending request is received by a forwarder. This can skew the client round-trip delay measurements posing another challenge for clients to accurately infer the current time. 
To address this challenge, \sol clients can randomize the name of their Interests, so that they avoid PIT aggregation. This can be achieved by attaching a random hash in the Interest name. Given that the \sol request name prefix is used for forwarding purposes, the randomization should happen in the request name suffix. We presented a mechanism to achieve name randomization in Section~\ref{sec:unicast}.

\section {Discussion}
\label{sec:disc}

{\abde In this section, based on the fundamental \sol design requirements and goals mentioned in Section~\ref{sec:design}, we highlight and discuss potential extensions of the \sol design as well as the different research directions that may be pursued as a continuation of this work.}

\subsection {In-network {\abde Time Synchronization}}



Given that NDN forwarders are aware of the communication context, they can identify {\spyros whether a specific Interest is an \sol request and whether} a specific Data packet is an \sol response. This can be particularly useful in the following cases: 
(i) {\spyros forwarders that} have an up-to-date estimation of the current time (\eg they recently received a timestamp from an accurate \sol server) can directly respond to \sol client requests with their own time. These responses are signed {\spyros by the forwarder itself}. This paradigm may constitute a distinct mode of operation that forwarders can enable/disable based on their load, available resources, and management policies; and (ii) {\spyros forwarders can utilize} ongoing exchanges between \sol clients and servers to satisfy their own time synchronization requirements. Specifically, forwarders can set their own clock based on the content of \sol responses forwarded by them back to clients.


\subsection {Strata Organization and Synchronization}

NTP servers are organized in strata that determine their distance from a reference clock. The larger the stratum number, the further away a server is from the reference clock. Servers that belong to stratum $N+1$ synchronize their clocks with servers that belong to stratum $N$, servers that belong to stratum $N$ synchronize their clocks with servers of stratum $N-1$, etc. NTP servers of the same stratum can also peer with each other to address {\spyros clock inconsistencies and achieve reliability}. To achieve this synchronization/peering process, \sol servers that belong to a specific stratum can announce a name prefix \name{/NDNTP/time/stratum=<stratum-number>}. For example, servers that belong to stratum 2 use the namespace \name{/NDNTP/time/stratum=2} for peering purposes with other servers of the same stratum. To synchronize their clocks with servers of stratum 1, they use the namespace \name{/NDNTP/time/stratum=1}.

\subsection{Security}
\label{subsec:security}

\sol utilizes NDN's network-layer security principles for response authentication purposes. Specifically, clients based on the signature of responses can: (i) verify {\spyros that the responses have} not been spoofed (\eg due to man-in-the-middle attacks); and (ii) decide whether they trust the servers that generated responses based on a pre-established set of trust anchors~\cite{zhang2018overview}. 

NDN also provides a solid foundation for the mitigation of DDoS attacks directly at the network layer. The NDN stateful forwarding plane can limit/throttle DDoS traffic close to its source(s) on a per name prefix basis. In packet delay attacks, adversaries delay time synchronization requests and responses between clients and servers in an asymmetric manner in order to skew the round-trip delays measured by clients~\cite{sibold2016network}. Such attacks can be mitigated through mechanisms for forwarding \sol requests over different network paths towards time servers. Furthermore, a threshold value for acceptable round-trip delays {\spyros between time synchronization requests and the corresponding responses} can be introduced--clients discard responses with round-trip delays larger {\spyros than the threshold.}

None of the mechanisms above protects against servers that turn malicious over time (\eg due to a security breach) or rogue forwarders that cache {\spyros server time samples in} order to satisfy client requests with outdated time samples. To address such cases, we can use a Distributed Ledger (DL) to log transactions (\ie timestamps received from servers) and verify if specific servers have sent inaccurate/bogus timestamps to clients. This DL can be implemented through {\spyros an NDN distributed dataset synchronization protocol}. Clients and a number of Verifier Nodes (VNs)--for example, a group of clients or trusted entities--{\spyros form a synchronization group}, 
so that transactions added to the DL are received by every party in the group. The verification process can be: (i) proactive, where VNs mine the DL to verify the transaction correctness; or (ii) reactive, where clients report inconsistent/inaccurate timestamps triggering the verification process (assuming that the majority of selected servers is legitimate). 



\section {Conclusion and Future Work}
\label{sec:concl}

In this paper, we presented the challenges of designing \sol, a time synchronization protocol that provides functionality analogous to NTP over NDN, and we discussed general {\spyros directions for its} design. Our concluding remarks indicate that the legacy NDN architectural design needs to be augmented to support the requirements of an NTP-like protocol, such as reaching multiple time servers at the same time and returning multiple responses (one from each time server) back to clients, requesting multiple time samples {\spyros from a set of selected servers}, and reaching time servers within certain distances from clients. Future directions include designing in detail, implementing, and evaluating an \sol prototype that takes advantage of the NDN feature enhancements proposed in this paper for accurate time synchronization. {\abde We also plan to extend this work by performing an analytical study to improve the server selection process using a multi-objective function}. 

\section*{Acknowledgements}
This work is partially supported by a pilot award from the Center for Research in Human Movement Variability and the NIH (P20GM109090), a planning award from the Collaboration Initiative of the University of Nebraska system, and the Nebraska Tobacco Settlement Biomedical Research Development Funds.

\bibliographystyle{IEEEtran}
\balance
\bibliography{refs}

\section*{Biographies}

\vskip -2.0\baselineskip plus -1fil

\begin{IEEEbiographynophoto}{Abderrahmen Mtibaa}
(amtibaa@umsl.edu) is currently an Assistant Professor at Department of Computer Science in the University of Missouri--St. Louis. Prior to that he has occupied several research positions including a visiting assistant professor at the Computer Science department in New Mexico State University; a Research Scientist at Texas A\&M University; and a Postdoc in the School of Computer Science at Carnegie Mellon University. His current research interests include Information-Centric Networking, Networked Systems, Social Computing, Personal Data, Privacy, IoT, mobile computing, pervasive systems, mobile security, and mobile opportunistic networks/DTN.
\end{IEEEbiographynophoto}

\vskip -2.0\baselineskip plus -1fil

\begin{IEEEbiographynophoto}{Spyridon Mastorakis}
(smastorakis@unomaha.edu) is an Assistant Professor in Computer Science at the University of Nebraska Omaha. He received his Ph.D. in Computer Science from the University of California, Los Angeles (UCLA) in 2019. He also received an MS in Computer Science from UCLA in 2017 and a 5-year diploma (equivalent to M.Eng.) in Electrical and Computer Engineering from the National Technical University of Athens (NTUA) in 2014. His research interests include network systems and protocols, Internet architectures (such as Information-Centric Networking and Named-Data Networking), and edge computing.
\end{IEEEbiographynophoto}

\end{document}